 \documentclass[final,3p,times,twocolumn]{elsarticle}

\usepackage{amssymb}
\usepackage{amsmath}
\usepackage{color}

\journal{Journal of Magnetism and Magnetic Materials}

\begin{document}

\begin{frontmatter}



\title{Strain Engineering of Spin and Rashba properties in Group-III Monochalcogenide MX (M=Ga, In and X=S, Se, Te) Monolayer}

\author{Mohammad Ariapour}
\address{Department of Sciences, Hamedan University of Technology, Hamedan 65155, Iran}
\author{Shoeib Babaee Touski}
\ead{touski@hut.ac.ir}
\address{Department of Electrical Engineering, Hamedan University of Technology, Hamedan 65155, Iran}

\begin{abstract}
In this paper, spin properties of monolayer MX (M=Ga, In and X=S, Se, Te) in the presence of strain are studied. Density functional theory is used to investigate spin properties. The strain changes modification of bandgap due to spin-orbit coupling, the results indicate the spin-orbit coupling has a higher effect in the compressive regime. Also, spin splitting in the conduction and valence bands respect to strain are compared for six materials. The location of conduction band minimum (CBM) imposed a type of spin properties. These materials with mirror symmetry can display the Rashba effect while M valley is located at CBM. Strain tunes the conduction band minimum in three valleys (K, M and $\Gamma$ valleys) and determines which spin effect (spin splitting, Rashba splitting or no spin splitting) has occurred in each strain for every material. Lastly, the relation between the Rashba parameter and the atomic mass is explored and it is observed that there is a linear correlation between atomic mass and Rashba coefficient. 
\end{abstract}


\end{frontmatter}


\section{introduction}
Since the last two decades, there has been a huge interest in  materials by strong spin-orbit interaction (SOI), which is very important for  understand relativistic quantum phenomena and has applications to spintronic devices. SOI contributes to known effects such as Rashba, spin-Hall effect, topological insulator, and Zeeman states \cite{ishizaka2011giant,sakano2013strongly,kim2015layered,liu2015crystal}. The recent studies show that relativistic spin–orbit interaction causes the spin polarization in non-magnetic materials by the local asymmetry (atomic site group) in non-centrosymmetric zinc-blende or wurtzite semiconductors\cite{riley2014direct,schaibley2014spintronics}.
In crystal structures without inversion symmetry, the spin splitting of the bands takes place called Dresselhaus effect and in 2D structures or surfaces, which is known as Rashba effect; however, these can be considered as different  representation of the same phenomenon \cite{zhang2014hidden}.

The effective operation of the electron spin in a semiconductor can come from the Rashba SOI due to the coupling of the linear momentum of the electron with its spin via the external electric field. Moreover, in the surface or interface, the inversion asymmetric potential further removes the spin degeneracy on 2D band structures \cite{rashba1960properties}. The principal feature of the Rashba effect is controlling the spin electronic states in the absence of a magnetic field in two-dimensional materials. The electron spin states split the perpendicular potential asymmetry with the spin–orbit interaction. However, it is hard to find a suitable 2D film with a large Rashba coefficient. Strong SOI and broken inversion symmetry are two important properties to have a great Rashba effect. In graphene and non-polar two-dimensional materials, such as transition metal monochalcogenides, breaking inversion symmetry is often arrived by applying out-of-plane electric fields or through inter-facial effects \cite{min2006intrinsic,yuan2013zeeman,avsar2014spin}.

\begin{figure*}
	\centering
	\includegraphics[width=1\linewidth]{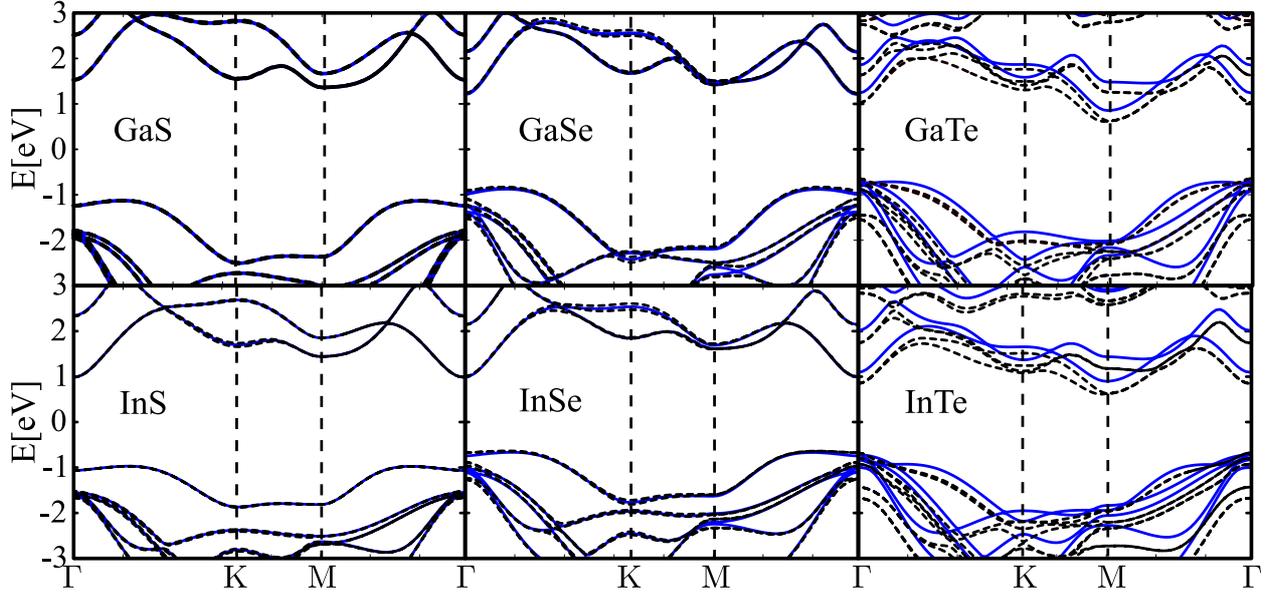}
	\caption{Band structure for MX (M=Ga, In, X=S, Se, Te) with (black dashed line) and without (blue solid line) SOC.}
	\label{fig:fig1}
\end{figure*}

In the last years, the manipulation of spin states in semiconductor devices has been a subject of attention of many papers
\cite{liu2006transport,ohno2008datta,nazmitdinov2009spin,thorgilsson2012rashba,sadreev2013effect}. 
The change of the spin direction in the spin-polarized current is a basic principle of the operation of proposed spintronic devices including the spin transistor \cite{datta1990electronic}. 
The inversion asymmetry has been an important impact in spintronics \cite{vzutic2004spintronics,nitta1997gate} as this induces a spin-orbit field. Datta-Das spin field-effect transistor \cite{datta1990electronic,chuang2015all} uses from the spin-orbit field, which the Rashba spin-orbit field controlled the coherent spin precession\cite{bychkov1984properties}.

Two-dimensional monolayer of group-III monochalcogenides \cite{huang2016superconductivity,zong2008enhancement} has attracted remarkable attention due to their various applications in spintronics, electronics \cite{yun2012thickness,radisavljevic2011v}, 
valleytronics\cite{suzuki2014valley} and photonics\cite{wang2015bottom}. Compounds with monolayer structures from Groups III-VI groups MX (M=Ga, In, X=S, Se, Te) with buckled hexagonal geometry were originated to have a strong Rashba effect and the band splitting\cite{ariapour19,di2015emergence,yao2017manipulation}. These features show the way for the design of pioneer spin field-effect transistors.

In this work, we explore the spin properties of monolayer crystals  MX (M=Ga, In,  X=S, Se, Te). The strain was used to tune spin properties of these materials\cite{jalilian17,ariapour19}. We investigate the
Rashba effect and spin-splitting in these monolayers  by applying  biaxial in-plane strain.

\begin{figure}[t]
	\includegraphics[width=1.0\linewidth]{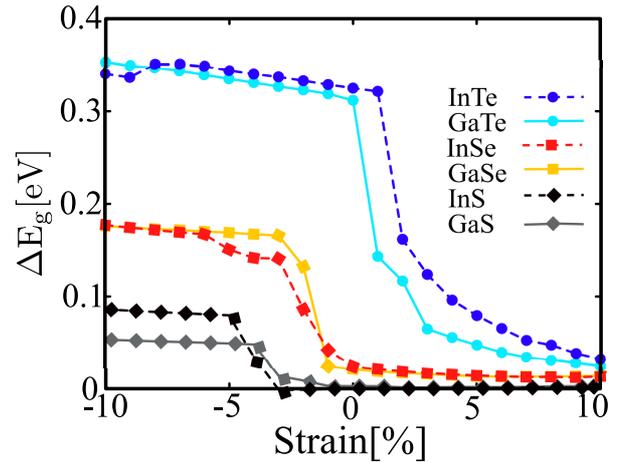}
	\caption{$\Delta E_g$ as a function of strain for the six materials. } 
	\label{fig:fig2}
\end{figure}

\section{Computational details}
Our results are based on density functional theory (DFT) calculations as implanted in the Quantum ESPRESSO package \cite{giannozzi2009quantum}. The general gradient approximation (GGA) of Perdew, Burke, and Ernzerhof (PBE) is adopted to describe the exchange-correlation potential \cite{perdew1996generalized}. We employed Projector Augmented-Wave (PAW) type pseudopotentials with a generalized gradient approximation using Perdew-Burke-Ernzerhof that revised for solids (PBEsol) functional \cite{perdew2008restoring}. The plane-wave cutoff energy was employed 612 eV for the plane-wave basis. All the structures are fully relaxed with a force tolerance of $0.001 eV/\AA$. Rashba calculation needs a dense k-point grid. During calculation, a $24\times24\times1$ Monkhorst-Pack k-point grid was used for Rashba calculations.

\section{results and discussion}

\begin{figure}[t]
	\includegraphics[width=0.9\linewidth]{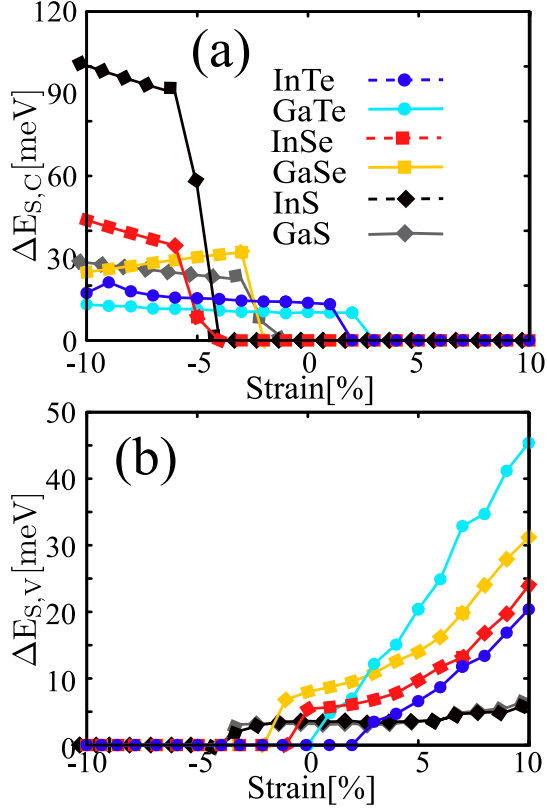}
	\caption{Spin Splitting of (a) conduction and (b) valence bands as a function of strain for MX materials.} 
	\label{fig:fig3}
\end{figure}

\begin{figure*}[t]
	\includegraphics[width=1\linewidth]{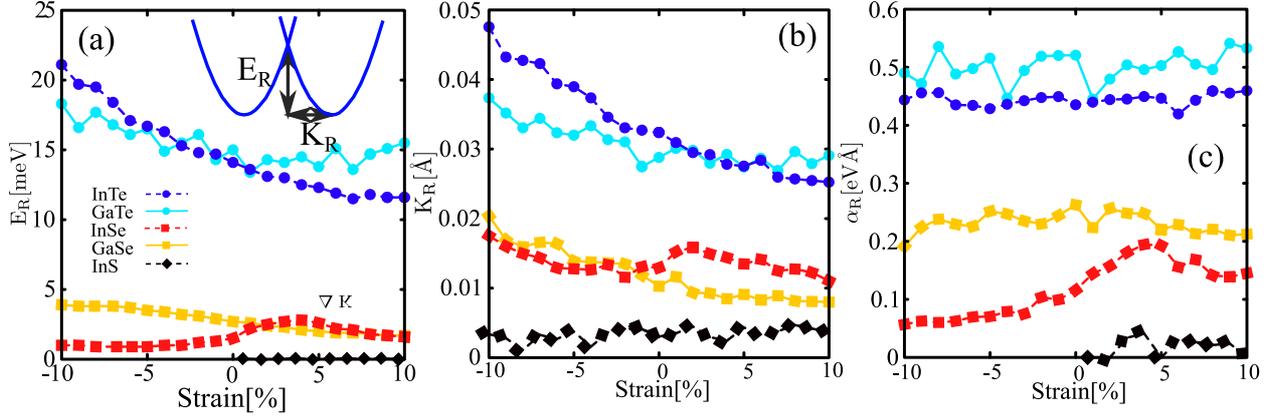}\\
	\caption{(a) Rashba energy as a function of energy. Inset figure shows defining of Rashba energy ($E_R$) and Rashba momentum ($K_R$). (b) Rashba momentum and (c) Rashba parameters versus strain for MX materials.} 
	\label{fig:fig4}
\end{figure*}

Band structures for six materials with and without spin-orbit coupling are shown in Fig. \ref{fig:fig1}. Band structure in GaS for with and without SOC are the same. This indicates SOC has a negligible effect on this material. InS behaves similar to GaS. The effects of SOC on the band structure increases ones changes  S to Se and Te. As one can observe, the band structures for InTe and GaTe is significantly different with SOC applied.

Band gaps for two states, with and without spin-orbit coupling, are calculated. The difference between these two band gaps, $\Delta E_{g}=E_g^{w/o}-E_g^{w}$, indicates the effects of SOC in these materials. $\Delta E_{g}$ is plotted respect to strain $((l-l_{0})/l_{0})$  in Fig. \ref{fig:fig2} where $l(l_{0})$ is the lattice constant under the strain (equilibrium) condition. The strain varies in the range  [$-10\%,10\%$] from equilibrium condition. Band gap decreases more than 0.3eV considering SOC for InTe and GaTe in compressive strain and 0.2eV for GaSe and InSe. In the compressive strain regime, we find that $\Delta E_{g}$ increases a little, whereas in the tensile strain this difference decreases respect to strain. $\Delta E_{g}$ behaves similarly for the compounds with similar chalcogenide atom (S, Se, Te). One can conclude spin properties are highly controlled with chalcogenide atoms. However, one can observe that $\Delta E_{g}$ is different for InS and GaS. $S$ atom is a light atom where In and Ga atoms with higher atomic mass control spin properties in these two compounds. With the change of chalcogenide atom (X) from S to Se and Te, $\Delta E_{g}$ increases for both compounds InX and GaX. This is due to the stronger SOC in heavier atoms. One can see that $\Delta E_g$ falls down for specific strain in every compound. In this strain, valley moves from K- and M-valley to $\Gamma$-valley.

The spin-splitting in the uppermost valence band and the lowermost conduction band in the presence of biaxial strain is plotted in Fig. \ref{fig:fig3}. Spin splitting in the conduction band ($\Delta E_{S,C}$) goes to zero for the tensile strain. The tensile strain moves valley to $\Gamma$-point and vanishes spin splitting. $\Delta E_{S,C}$ increases with compressive strains. The spin-splitting in valence band goes to zero for compressive strain. For compressive strain, Mexican hat shape of valence band is vanished and top of valence band located at $\Gamma$-point and spin splitting vanishes. Moving from compressive strain to tensile strain, Mexican hat is created at top of valence band and the top of valence band goes to $\Gamma^*$- from $\Gamma$-point \cite{ariapour19}. In tensile strain, spin splitting in the valence band ($\Delta E_{S,V}$) increases with the increase of tensile strain.

The Group-III monochalcogenides show Rashba spin splitting at M-point \cite{ariapour19}. Rashba splitting is characterized with three parameters, Rashba energy ($E_{R}$), Rashba momentum ($K_{R}$), and Rashba coefficient ($\alpha_{R}$). Rashba energy and momentum are calculated, see inset on Fig. \ref{fig:fig4}(c) and $\alpha_{R}$ is defined as: $\alpha_{R}=2 E_R/K_R$. These parameters are investigated for these materials and the results are plotted in Fig. \ref{fig:fig4}. Rashba energy and Rashba momentum rises with the increasing strain in the compressive strain, whereas, they decrease by increasing strain in the tensile regime. $E_R$ for InS is very small, close to zero and we didn't observe any Rashba spin splitting in GaS.  $K_R$ also declines with the increasing strain. The heavier compounds show higher $K_R$ like $E_R$. Rashba coefficient as a function of strain is plotted for five compounds, see Fig. \ref{fig:fig4}(c). Rashba coefficient remains constant with strain for InTe, GaTe, and GaSe compounds. However, $\alpha_{R}$ for InSe increases with the increasing strain and decreases for strain larger than 5$\%$. We didn't observe any Rashba spin splitting in InS in compressive strain whereas tensile strain exhibits a weak Rashba splitting. We expect heavier compounds exhibit higher $\alpha_{R}$ Whereas one can see, InTe is heavier than GaTe; GaTe shows the highest Rashba coefficient. Such effect happens for InSe and GaSe, too. GaSe with lighter atoms exhibits higher $\alpha_{R}$ relative to GaTe.

\begin{figure}[t]
	\includegraphics[width=0.99\linewidth]{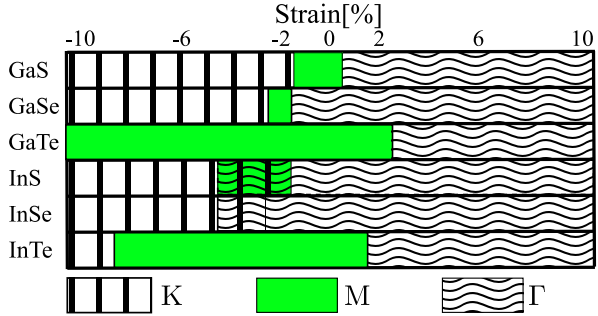}
	\caption{Minimum of conduction band is discussed for different compounds at various strain. Blue color exhibits CBM is located at K and amount of spin splitting in this point is shown with blue color. Red color shows CBM is located at M point and compound is a Rashba semiconductor. Blank place exhibit CBM is located at $\Gamma$-point and there is not any spin splitting.   } 
	\label{fig:fig5}
\end{figure}

Spin splitting at K-valley and Rashba splitting at M-valley was discussed, however, conduction band minimum (CBM) is located at one valley for a specific strain. Therefore, the location of CBM between valleys for six materials in the presence of biaxial strain is reported in Fig. \ref{fig:fig5}. CBM is located at K-valley for large compressive strain and moves to M-valley when strain increases, and then moves to $\Gamma$-valley for large tensile strain. For example, CBM in GaS is located at K-valley for strain smaller than -2$\%$ and this material exhibits spin splitting in this strain range. CBM at M-valley for strain [-2$\%$-0$\%$] causes this material to behave as a Rashba semiconductor. For strain larger than 0$\%$, the place of CBM is at $\Gamma$-valley and $\Gamma$-valley doesn't indicate any spin-splitting due to high symmetry of $\Gamma$-point. Both InTe and GaTe exhibit large Rashba coefficient relative to other materials, see Fig. \ref{fig:fig4}, and CBM is located at M-valley in a wide range of strain for these two materials. InTe is explored in our previous work in which the location of CBM is compatible with our previous work \cite{touski2019electrical}. The location of CBM is not placed at K-valley in GaTe for any strain so this material doesn't show spin splitting. CBM is not located at M-valley for any range of strain in InSe, so this material doesn't show any Rashba properties. One can see for strain $-3\%$ in InSe, that CBM is placed in two K- and $\Gamma$-valleys. In addition, the CBM is located at K, M and $\Gamma$-valleys for strain in the interval [-4$\%$,-2$\%$].

\begin{figure*}
	\includegraphics[width=0.8\linewidth]{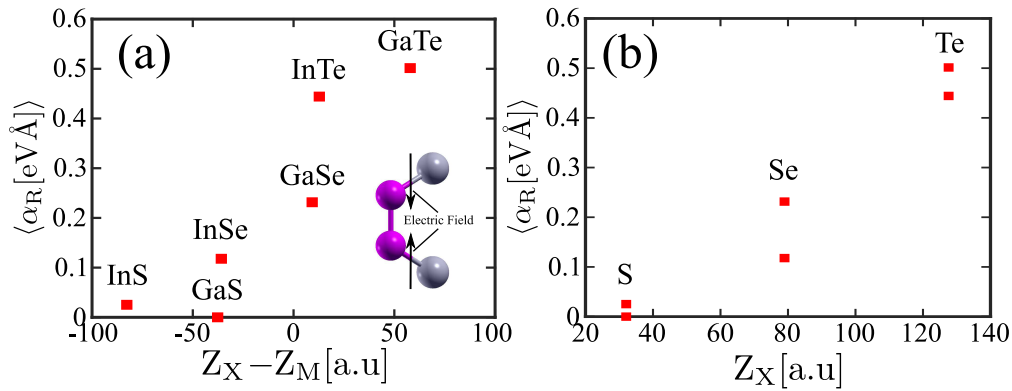}
	\caption{(a) Rashba coefficient is plotted versus the atomic mass of M minus the atomic mass of X (m$_M$-m$_X$). The inset figure shows a unit cell of MX with an effective electric field. (b) Rashba coefficient versus the atomic mass of chalcogenide.} 
	\label{fig:fig6}
\end{figure*}

In systems with a similar structure, the different values of the Rashba coefficient can be attributed to the difference in atomic mass\cite{gierz2010structural}. For example, Rashba coefficient for Bi/Ag is reported 3.05eV$\AA$ \cite{ast2007giant} as much as 0.76 eV$\AA$ for Sb/Ag \cite{moreschini2009assessing,meier2011interference}.  The atomic mass of Bi is twice of Sb and $\alpha_{R}$ in Bi is four-time larger than SB. This indicates Rashba splitting is proportional to $Z^2$. However, we didn't observe such dependency for Rashba coefficient. To clarify the behavior of Rashba coefficient, the average of $\alpha_{R}$ is plotted as a function of atomic mass of X ($Z_X$) minus the atomic mass of M ($m_M$), see Fig. \ref{fig:fig6}. An electric field is created due to the difference between the potential of two atoms in MX sub-layer. This electric field causes an effective magnetic field that results in Rashba property. The difference between the atomic mass of chalcogenide and group-III atoms, $\Delta Z=Z_X-Z_M$, influences the electric field. One can observe that approximately there is a linear relation between $\Delta Z$ and Rashba coefficient. $\Delta Z$ is the highest for GaTe where this compound shows the highest $\alpha_{R}$, whereas In has a larger atomic mass and InTe should indicate the highest Rashba coefficient. In addition, the Rashba coefficient is plotted as a function of chalcogenides atomic mass. This coefficient is close to each other for compounds with similar chalcogenide atom. This indicates the Rashba coefficients are highly affected by chalcogenide atoms.

\section{conclusions}
The effects of strain on the spin properties of group III-monochalcogenides were studied. The results show spin-orbit coupling decreases the band gap and is more effective for heavy materials. The SOC declines the band gap more than 0.3eV for InTe and GaTe with the compressive strain. Spin-splitting in the conduction band increases with the increasing compressive strain, but in the valence band with the tensile strain. Rashba splitting parameters are obtained for all materials and Rashba momentum and Rashba energy decreases with the increasing strain, whereas Rashba coefficient remains constant. The heavier compounds show higher Rashba parameters where GaTe and InTe show the highest Rashba coefficient close to 0.5 eV$\AA$. We find the Rashba parameter highly depends on X (S, Se, and Te) atoms, showing a correlation with the difference between the mass of M and X atoms.  



\end{document}